\documentclass[12pt, a4paper]{article}
\usepackage{a4wide,amssymb,amsmath,amscd}

\begin{document}

\topmargin -2pt


\headheight 0pt

\topskip 0mm \addtolength{\baselineskip}{0.20\baselineskip}
\begin{flushright}
{\tt KIAS-P07040}
\end{flushright}

\vspace{5mm}

\begin{center}
{\Large \bf Charged Black Holes on DGP Brane} \\

\vspace{10mm}

{\sc Ee Chang-Young}${}^{ \dag,
\ddag, }$\footnote{cylee@sejong.ac.kr} ~ and ~ {\sc Daeho Lee}${}^{  \dag,  }$\footnote{dhlee@sju.ac.kr}\\

\vspace{1mm}

 ${}^{\dag}${\it Department of Physics, Sejong University, Seoul 143-747, Korea}\\

${}^{\ddag}${\it School of Physics, Korea Institute for Advanced
Study,
Seoul 130-722, Korea}\\

\vspace{10mm}

{\bf ABSTRACT} \\
\end{center}


\noindent
We consider charged black holes localized on a
three-brane in the DGP model. With a spherically symmetric metric
ansatz on the brane and assuming a $Z_2$-symmetry across the
brane, we find two types of solutions which correspond to the
so-called regular and accelerated branches in the DGP model. When
the charge of the black hole vanishes, our solutions agree with
the Schwarzschild solutions obtained by Gabadadze and Iglesias.
\\

\vfill


\thispagestyle{empty}

\newpage




%

\section{Introduction}

Recent astronomical observations indicate that our universe is in
the phase of accelerated expansion \cite{sauv}. The DGP model
\cite{dgp} is known to contain a branch of solutions exhibiting
self accelerated expansion of the universe \cite{df}.

 The brane-world black holes in the case of Randall-Sundrum(RS) model \cite{RS12}
 were  first considered in \cite{chr:prd}.
 Then, Dadhich et al.\cite{dmpr:plb} showed that for static black holes localized
 on a three-brane in the RS model the Reissner-N\"{o}rdstrom metric is an
 exact solution even when black holes are uncharged.
 They interpreted it as a `tidal charge' effect arising via gravitational effect
 from the extra dimension.
Then a solution of charged brane-world black holes in the RS model
was obtained in \cite{crss:prd}, and the charged rotating case was
obtained in \cite{ag:prd}.
On the other hand, for the DGP model only approximate
Schwarzschild solutions had been obtained in \cite{po,kp,ls,ms,nr}
until Gabadadze and Iglesias obtain the exact Schwarzschild
solutions in \cite{gi:prd}.

 However, in \cite{kol:prd} the instability issue of the static uncharged black
holes on the brane was raised.
The brane with the
induced graviton kinetic term has effectively repulsive gravity
and thus push any sources off the brane.
It was argued that the charged black holes could be quasilocalized when the
corresponding gauge fields are localized.  Thus the need of investigating the charged case
was called for in \cite{gi:prd}.
 So far, the solutions of charged or rotating black holes on the
 brane in the DGP model have not been known.

In this paper, we investigate the solutions of charged black holes
located on the three-brane in the DGP model. We solve this by
assuming a spherically symmetric metric ansatz on the brane, and
find two types of solutions one of which exhibits the self
acceleration of the brane-world universe.

This paper is organized as follows. In section II, we set the
action and equations of motion of the DGP model following the
approach of Ref. \cite{ag:cqg}. In section III, we give exact
solutions for the metric on the brane in the presence of the Maxwell
field. In section IV, we conclude with the discussion.
\\

\section{Action and field equations}

The DGP gravitational action in the presence of sources takes the form \cite{dgp}
\begin{eqnarray}
\label{action}
S = M_{*}^{3}\int d^5 x \sqrt{-g}~^{(5)}R + \int d^4 x \sqrt{-h} \left(M_{P}^{2}R
+L_{matter}\right),
\end{eqnarray}
where $R$ and $^{(5)}R$ are the 4D and 5D Ricci scalars,
respectively and $L_{matter}$ is the Lagrangian of the matter
fields trapped on the brane. Here, the $(4+1)$ coordinates are
$x^{A}=(x^{\mu},y(= x^5))$, $\mu=0,1,2,3$, and $g$ and $^{(5)}R$
are the determinant and curvature of the five-dimensional metric
$g_{AB}$, while $h$ and $R$ are the determinant and curvature of
the four-dimensional metric $h_{\mu\nu}=g_{\mu\nu}(x^{\mu},y=0)$.
A cross-over scale is defined by $r_{c}=
m_{c}^{-1}=M_{P}^{2}/2M_{*}^{3}$.
There is a boundary(a brane) at $y=0$ and $Z_{2}$ symmetry across
the boundary is assumed. The boundary Gibbons-Hawking term
\cite{gh} is implied to yield the correct Einstein equations in
the bulk.
 The field equations derived from the action (\ref{action}) have the form
\begin{equation}
\label{eom:munu}
^{(5)}G_{AB}= ^{(5)}R_{AB}-\frac{1}{2}g_{AB}~^{(5)}R=
\kappa_{5}^{2}\sqrt{\frac{h}{g}}\left(X_{AB}+T_{AB}\right)\delta(y),
\end{equation}
where $\kappa_{4}^{2}=M_{P}^{-2}$ and $\kappa_{5}^{2}=M_{*}^{-3}$,
while $
X_{AB}=-\delta_{A}^{\mu}\delta_{B}^{\nu}G_{\mu\nu}/\kappa_{4}^{2}$
and $T_{AB}= \delta_{A}^{\mu}\delta_{B}^{\nu}T_{\mu\nu}$ is the
energy-momentum tensor in the braneworld.
 Now, we consider the metric of the following form,
\begin{equation}
\label{metric}
ds^2=g_{AB}dx^{A}dx^{B}=g_{\mu\nu}(x,y)dx^{\mu}dx^{\nu}+2N_{\mu}dx^{\mu}dy+(N^2
+g_{\mu\nu}N^{\mu}N^{\nu})dy^2,
\end{equation}
where the so-called lapse scalar field $N$ and the shift vector
field $N_{\mu}$ are defined by
 \cite{ag:cqg,dgl:jcap}
\begin{equation}
\label{NNmu}
g_{\mu 5} \equiv N_{\mu}=g_{\mu\nu}N^{\nu},~g_{55} \equiv N^{2}+g_{\mu\nu}N^{\mu}
N^{\nu}.
\end{equation}

The $(\mu 5)$, $(55)$ components of the field equations
(\ref{eom:munu}) are called as the momentum and Hamiltonian
constraint equations, respectively, and are given by
\cite{ag:cqg,dgl:jcap}
\begin{equation}
\label{eom:5i}
\nabla_{\nu}K^{\nu}_{~\mu}-\nabla_{\mu}K=0,
\end{equation}
\begin{equation}
\label{eom:55}
R-K^{2}+K_{\mu\nu}K^{\mu\nu}=0,
\end{equation}
where $K_{\mu\nu}$ is the extrinsic curvature tensor defined by
\begin{equation}
\label{kmunu}
K_{\mu\nu}=\frac{1}{2N}(\partial_{y} g_{\mu\nu}-\nabla_{\mu}N_{\nu}-\nabla_{\nu}
N_{\mu}),
\end{equation}
and $\nabla_{\mu}$ is the covariant derivative operator associated with the metric
$g_{\mu\nu}$.

To implement the Israel's junction condition \cite{Israel}, we
integrate both sides of the field equation (\ref{eom:munu}) along
the $y$ direction and take the limit of $y=0$ on the both sides of
the brane:
\begin{equation}
\lim_{\epsilon \rightarrow 0}\int_{-\epsilon}^{+\epsilon} dy ~^{(5)}G_{\mu\nu}=
\lim_{\epsilon \rightarrow 0}\int_{-\epsilon}^{+\epsilon} dy~\kappa_{5}^{2}
\sqrt{\frac{h}{g}}(X_{\mu\nu}+T_{\mu\nu})\delta(y),
\end{equation}
where
\begin{eqnarray}
\label{5Gmunu}
^{(5)}G_{\mu\nu} &=& G_{\mu\nu}-\frac{1}{N}[(\partial_{y}-\mathcal{L}_{\vec{N}})
(K_{\mu\nu}- g_{\mu\nu}K)+\nabla_{\mu}\nabla_{\nu}N] \nonumber \\
&&
-3KK_{\mu\nu}+2K_{\mu}^{~\rho}K_{\rho\nu}+\frac{1}{2}g_{\mu\nu}(K^2+
K_{\rho\sigma}K^{\rho\sigma}+\frac{2}{N}\nabla_{\rho}\nabla^{\rho}
N).
\end{eqnarray}
Here the Lie derivative $\mathcal{L}_{\vec{N}}$ is taken along the
shift vector $N^{\mu}$, and the Israel's junction  condition
becomes,
\begin{equation}
\label{pprejc}
\lim_{\epsilon \rightarrow 0}\left[ -\frac{1}{N} (K_{\mu\nu}-g_{\mu\nu}K)
\right]_{-\epsilon}^{+ \epsilon}= \kappa_{5}^{2} \sqrt{\frac{h}{g}}(X_{\mu\nu}+
T_{\mu\nu}).
\end{equation}
Using the relation $\frac{1}{N}=\sqrt{\frac{h}{g}}$, this can be
written as
\begin{equation}
\label{prejc}
[K_{\mu\nu}] - h_{\mu\nu}[K]=-\kappa_{5}^{2} (X_{\mu\nu}+T_{\mu\nu}),
\end{equation}
where $[A] \equiv A|_{y=0^{+}}-A|_{y=0^{-}}$.
 Imposing the $Z_{2}$-symmetry across the brane,
\begin{equation}
K_{\mu\nu}^{+}=-K_{\mu\nu}^{-},
\end{equation}
and taking the limit $y=0$, the junction condition (\ref{prejc})
can be expressed as
\begin{equation}
\label{jc}
G_{\mu\nu}=\kappa_{4}^2 T_{\mu\nu}+m_{c}(K_{\mu\nu}-h_{\mu\nu}K).
\end{equation}

Once we take the electro-magnetic field as the matter source on
the brane, the trace of the energy momentum tensor for the
electro-magnetic field vanishes. The momentum constraint equation
(\ref{eom:5i}) on the brane is  satisfied identically due to
(\ref{jc}),
\begin{eqnarray}
\label{momentumc}
0 &=& D_{\mu}K_{~\nu}^{\mu}-D_{\nu}K =D^{\mu}(K_{\mu\nu}-h_{\mu\nu}K),
\end{eqnarray}
where $D_{\mu}$ is the covariant derivative operator associated
with the brane metric $h_{\mu\nu}$.
 The Hamiltonian constraint on the brane (\ref{eom:55})
 is given by
\begin{eqnarray}
\label{Hamiltonc}
 0  &=& R+\frac{1}{m_c^2}(\kappa_{4}^{4}
T_{\mu\nu}T^{\mu\nu}+R_{\mu\nu}
  R^{\mu\nu}-\frac{1}{3}R^{2}-2\kappa_{4}^2 R_{\mu\nu}T^{\mu\nu}).
\end{eqnarray}
In what follows we shall set $\kappa_{4}^2=8\pi$.

Finally, combining the five-dimensional Einstein Equations
\begin{equation}
^{(5)}G_{AB}= ^{(5)}R_{AB}-\frac{1}{2}g_{AB}~^{(5)}R=0,
\end{equation}
with the Israel's junction condition (\ref{jc}), we arrive at the
gravitational field equations on the brane
\begin{equation}
\label{effge}
G_{\mu\nu}=-E_{\mu\nu}-\kappa_{5}^{4}\tilde{\tau}_{\mu\nu},
\end{equation}
where $E_{\mu\nu}$, the traceless ``eletric part" of the
5-dimensional Weyl tensor $^{(5)}\!C_{ABCD}$, is defined by
\cite{sms:prd}
\begin{equation}
\label{Amunu}
E_{\mu\nu}= ^{(5)}\!C_{ABCD}n^{A}n^{C}e_{\mu}^{~B}e_{\nu}^{~D},
\end{equation}
and
\begin{equation}
\label{tildetaumunu}
\tilde{\tau}_{\mu\nu}=\frac{1}{4}[ \tau^{\rho}_{~\mu}\tau_{\rho\nu}
-\frac{1}{3}\tau\tau_{\mu\nu}
-\frac{1}{2}h_{\mu\nu}(\tau_{\rho\sigma}\tau^{\rho\sigma}
-\frac{1}{3}\tau^{2})],
\end{equation}
with $\tau_{\mu\nu}=T_{\mu\nu}+X_{\mu\nu}$.
It should be noted that $E_{\mu\nu}$ above is the limiting value
at $y=+0$ or $-0$ but not the value exactly on the brane. Using
the relation $X_{\mu\nu}=-G_{\mu\nu}/\kappa_{4}^{2}$, the
gravitational field equation (\ref{effge}) on the brane is given
by
\begin{eqnarray}
\label{grave}
G_{\mu\nu} &=&-E_{\mu\nu}-\frac{\kappa_{4}^{4}}{m_{c}^2}(T^{\rho}_{~\mu}T_{\rho\nu}
-\frac{1}{2}h_{\mu\nu}T_{\rho\sigma}T^{\rho\sigma})
     \nonumber \\
&& -\frac{1}{m_{c}^2}(R^{\rho}_{~\mu}R_{\rho\nu}-\frac{2}{3}RR_{\mu\nu}
+\frac{1}{4}h_{\mu\nu}R^2
-\frac{1}{2}h_{\mu\nu}R_{\rho\sigma}R^{\rho\sigma})  \nonumber \\
&& +\frac{\kappa_{4}^2}{m_{c}^2}(R^{\rho}_{~\mu}T_{\rho\nu}
+T^{\rho}_{~\mu}R_{\rho\nu}-\frac{2}{3}RT_{\mu\nu}
           -h_{\mu\nu}R_{\rho\sigma}T^{\rho\sigma}),
\end{eqnarray}
where we used $T=T^{\mu}_{~\mu}=0$ and $m_{c}^{-1}=\kappa_{5}^2 /2
\kappa_{4}^2$.
If we take the trace of (\ref{grave}), we get the Hamiltonian
constraint (\ref{Hamiltonc}). Note that the
four-dimensional gravitational field equations (\ref{grave})
contains a term, $E_{\mu\nu}$, which carries information of the
gravitational field outside the brane.
In general, the field equations on the brane are not closed and
one needs to solve the evolution equations into the bulk. However,
by assuming a special ansatz for the induced metric
 on the brane, one can make the system of equations on the brane closed.

\section{Charged black hole solutions}

 For charged black holes on the brane, we assume the following
spherically symmetric metric ansatz,
\begin{eqnarray}
\label{metricb}
ds^{2}&=& h_{\mu\nu}dx^{\mu}dx^{\nu} \nonumber \\
      &=& -\left(1-\frac{P(r)}{r}\right)dt^{2}+\left(1-\frac{P(r)}{r}\right)^{-1}dr^{2}
      +r^2 d\theta^{2}+r^2 \sin^{2}\theta d\phi^{2}.
\end{eqnarray}

The Hamiltonian constraint (\ref{Hamiltonc}) in the presence of a
Maxwell field can be written as
\begin{eqnarray}
\label{Hamiltoncmaxwell}
0=R_{\mu\nu}R^{\mu\nu}-\frac{1}{3}R^2+m_{c}^2 R-2\kappa_{4}^2 R_{\mu\nu}T^{\mu\nu}
+4\kappa_{4}^4 S,
\end{eqnarray}
where $S \equiv T_{\mu\nu}T^{\mu\nu}/4=Q^4/64\pi^2 r^8$,~
$T_{\mu\nu}=(F_{\mu\rho}F^{\rho}_{~\nu}-\frac{1}
{4}F_{\rho\sigma}F^{\rho\sigma})/4\pi$.

We assume that the Maxwell field on the brane is described by a
solution of source-free Maxwell equations. In this case, we must
solve both the Hamiltonian constraint (\ref{Hamiltoncmaxwell}) and
the Maxwell's equations:
\begin{equation}
\label{fMaxwell}
g^{\mu\nu}D_{\mu}F_{\nu\sigma}=0,
\end{equation}
\begin{equation}
\label{sMaxwell}
D_{[\mu}F_{\nu\sigma]}=0.
\end{equation}
Eq.(\ref{sMaxwell}) is satisfied identically if the field strength
$F_{\mu\nu}$ is constructed by the potential one-form. One can
easily verify that the solution of the Maxwell equations written
in terms of a potential one-form under the metric ansatz
(\ref{metricb}) has the following  form
\begin{equation}
\label{potential}
A=-\frac{Q }{r}dt,
\end{equation}
where the parameter Q is thought of as the electric charge of the
black hole.
The energy-momentum tensor for the above potential one-form $A$ is
calculated to be
\begin{eqnarray}
\label{em}
 T_{tt}=\frac{Q^2}{8\pi r^4}(1-P/r), ~ T_{rr}=-\frac{Q^2}{8\pi
 r^4}(1-P/r)^{-1},~
 T_{\theta\theta}=\frac{Q^2}{8\pi r^2},
 ~ T_{\phi\phi}=\frac{Q^2 \sin^2\theta}{8\pi r^2}.
\end{eqnarray}

Before solving the Hamiltonian constraint
(\ref{Hamiltoncmaxwell}), we consider particular solutions of this
constraint equation as a warm-up exercise. First, we notice that
the following set of relations satisfy the constraint equation (\ref{Hamiltoncmaxwell}):
\begin{equation}
\label{relation} R=12m_{c}^2,~
R_{\mu\nu}R^{\mu\nu}=36m_{c}^4+4Q^4/r^8,~
R_{\mu\nu}T^{\mu\nu}=Q^4/2\pi r^8.
\end{equation}
These relations are satisfied, if the function $P$ in the metric
ansatz (\ref{metricb}) is given by
\begin{equation}
\label{scalarP}
P=-\frac{Q^2}{r}+m_{c}^2r^3+C,
\end{equation}
where the parameter $C$ is an integration constant.

Now, we consider a general solution satisfying the Hamiltonian
constraint equation with the metric ansatz (\ref{metricb}).
With the energy-momentum tensor (\ref{em}) and the metric ansatz,
the Hamiltonian constraint (\ref{Hamiltoncmaxwell}) can be
expressed as
\begin{eqnarray}
\label{Hamiltonce}
0&=&6m_{c}^2 r^2 (2Z+rZ_{r})+r^2Z_{r}^2 -8rZ Z_{r}+4Z^2-\frac{12 Q^2}{r^2}(2Z-rZ_{r})
+\frac{24Q^4}{r^4}
\end{eqnarray}
where $Z=dP/dr$ and $Z_{r}=d^2 P/dr^2$.
 For $Q=0$, as one should expect, the above equation reduces to the equation
for the Schwarzschild case \cite{gi:prd}.
To learn the behavior of exact solution, we first study asymptotic
behaviors in two regions $r \gg r_c$ and $r \ll r_c$.

 For $r \gg
r_c$, we can neglect the last five terms on the right-hand side of
(\ref{Hamiltonce}) since the quadratic terms is viable only for
lage distances. The solution in this case is given by
\begin{eqnarray}
\label{Plarge}
P=\frac{\tilde{r}_{M}^2}{r}+C_{1}~~\texttt{for $r \gg r_{c}$},
\end{eqnarray}
where $\tilde{r}_{M}$ and $C_{1}$ are integration constants.
When $C_{1}=0$, this solution becomes the ordinary
five-dimensional Schwarzschild solution of radius $\tilde{r}_{M}$.

 For $r \ll r_c$,  the first term on the
right-hand side of (\ref{Hamiltonce}) can be neglected and reduces
to
\begin{eqnarray}
\label{r<<rc}
r^2Z_{r}^2 -8rZ Z_{r}+4Z^2-\frac{12 Q^2}{r^2}(2Z-rZ_{r})+\frac{24Q^4}{r^4}\approx 0.
\end{eqnarray}
Solving for $Z_{r}$, we get
\begin{eqnarray}
\label{Zrsmall}
Z_{r}=-\frac{1}{r}\left[ \frac{6Q^2}{r^2}-4Z \pm 2\sqrt{3}\left(\frac{Q^2}{r^2}-Z\right)\right].
\end{eqnarray}
 The solution for this is given by
\begin{eqnarray}
\label{Pinr<<rc}
P=r_{M}-\frac{Q^2}{r}+C_{1}r^{5 \pm 2\sqrt{3}}~~\texttt{for $r \ll r_{c}$},
\end{eqnarray}
where $r_{M}$ and $C_{1}$ is integration constants.
When $C_{1}=0$, this becomes the charged black hole solution for
mass $r_M/2$ and charge $Q$ in the general relativity.\\

 For an interpolating solution between these two regimes, we first solve
(\ref{Hamiltonce}) for $Z_{r}$ :
\begin{eqnarray}
\label{Zr}
Z_{r}=-\frac{1}{r}\left[\left(\frac{6Q^2}{r^2}-4Z\right)+3m_{c}^2 r^2\left(1\pm \sqrt{1+4u
+\frac{4}{3}u^2}\right)\right],
\end{eqnarray}
where  $u=(Q^2-r^2 Z)/m_{c}^2 r^4$.
Introducing a new function $U$ replacing $Z$,
\begin{equation}
\label{randZ}  Z=Q^2/r^2-3m_{c}^2 r_{0}^2 e^{2z}U(z)/2,
\end{equation}
with $r=r_{0}e^{z}$, Eq.(\ref{Zr}) can be reexpressed as
\begin{eqnarray}
\label{lr}
 -\frac{1}{r}\left[ \frac{2Q^{2}}{r^2}+\frac{3m_{c}^{2}r^{2}}{2}(2U+U_{z}) \right]
 =
 -\frac{1}{r}\left[ \frac{2Q^{2}}{r^2}+\frac{3m_{c}^{2}r^{2}}{2}(4U+2(1\pm f))\right],
\end{eqnarray}
where $f=\sqrt{1+6U+3U^2}$.
One can now solve this equation in terms of $U_{z}$;
\begin{equation}
\label{Ueq}
U_{z}=2(1+U \pm f).
\end{equation}

 Although the constraint equation (\ref{Hamiltonce}) is different from the Schwarzschild
case \cite{gi:prd}, we got the same equation (\ref{Ueq}) to solve. This equation has been
solved implicitly and it has two solutions \cite{gi:prd},
\begin{eqnarray}
\label{regulU}
 \ln \left[-\frac{(1+3U+f)}{F^2(3+3U+\sqrt{3}f)^{2\sqrt{3}}(-5-3U+f)}
  \right] &=& 8z+C_{3}
\end{eqnarray}
for the minus($-$) sign in the last term in (\ref{Ueq}), and
\begin{eqnarray}
\label{accelU}
 \ln \left[-\frac{(-5-3U+f)(-3-3U-\sqrt{3}f)^{2\sqrt{3}}}{(U+2)^2(1+3U+f)}
  \right]&=& 8z+C_{3}
\end{eqnarray}
for the plus($+$) sign, and $C_{3}$ is an integration constant.

Now, from $Z=dP/dr$ and (\ref{randZ}), $P(r)$ can be determined by
the following relation
\begin{eqnarray}
\label{P}
 P(r)=-\frac{Q^2}{r}-\frac{3}{2}m_{c}^{2} \int dr ~r^{2} U(r)+C,
\end{eqnarray}
where $C$ is an integration constant to be determined by the
boundary condition.
Note that Eq.(\ref{Ueq}) is satisfied with $U=-2$ for the plus
sign  and $U=0$ for the minus sign in (\ref{Ueq}), and the above relation for $P$ gives
the same particular solution that we obtained in (\ref{scalarP}).

 Finally, using the relation $r=r_{0}e^{z}$, we can reexpress
(\ref{regulU}) and (\ref{accelU}) in terms of $r$ as follows \cite{gi:plb},
\begin{eqnarray}
(k_{1}r)^8\!\!&=&\!\!
     -\frac{(1 + 3U + f)}{U^2 (3 + 3U +
       \sqrt{3}f)^{2\sqrt{3}}(- 5 - 3U + f)},
       \label{ktoRegulU} \\
 (k_{2}r)^8 \!\!&=&\!\!
    - \frac{(- 5 - 3U + f)(-3 - 3U -
       \sqrt{3}f)^{2\sqrt{3}}}{(U + 2)^2( 1 + 3U + f)}
        \label{ktoAccelU},
\end{eqnarray}
where the two integration constant $k_{1}$ and $k_{2}$ are
determined by imposing appropriate boundary conditions. The above
two solutions are the so-called regular and accelerated
branches, respectively \cite{gi:prd}.


For the regular branch solution (\ref{ktoRegulU}) we impose the boundary conditions, $P(r)+Q^2/r
\rightarrow r_{M}$ for $r \rightarrow 0$ and $P(r) \rightarrow 0$
for $r \rightarrow \infty$.
The integration constant $k_{1}$ can be obtained from (\ref{P}) and is given by
\cite{gi:plb}
\begin{equation}
\label{k1} 2(k_{1}r_{*})^3 \equiv c \approx 0.43,
\end{equation}
where $r_{*}=(r_{M}r_{c}^2)^{1/3}$. Here c is given by the following integral
\begin{equation}
c=\int_{0}^{\infty} dU \left[-\frac{(1 + 3U + f)}{U^2 (3 + 3U +
       \sqrt{3}f)^{2\sqrt{3}}(- 5 - 3U + f)}\right]^{3/8} .
\end{equation}

 We can easily see the following asymptotic behavior by consulting the result of \cite{gi:plb}.
At large distances, $r \gg r_{*}(U \rightarrow 0^{+})$, (\ref{ktoRegulU}) is approximated as
\begin{equation}
\label{lregulapproxU}
U \approx \frac{\sqrt{2}}{2 (3+\sqrt{3})^{\sqrt{3}}(k_{1}r)^4}.
\end{equation}
Using (\ref{k1}) and (\ref{lregulapproxU}) one obtains
\begin{equation}
\frac{P(r)}{r}\approx \frac{\tilde{r}^2_{M_{1}}-Q^2}{r^2},
\end{equation}
where $\tilde{r}^2_{M_{1}}=\frac{3\sqrt{2}}{4(3+\sqrt{3})^{\sqrt{3}}}\frac{m_{c}^2}{k_{1}^4}
\approx 0.56 r_{M}r_{*}$.

In the regular branch solution, there appeared the screened mass effect in the Schwarzschild
case \cite{gi:prd,gi:plb}. The same effect appears here, too.
However, there is no such effect for the charge, as one can expect from the form of our solution (\ref{P}).

 At short distances, $r \ll r_{*}(U \rightarrow +\infty)$, (\ref{ktoRegulU}) is approximated as
\begin{equation}
\label{sregulapproxU}
U \approx 6^{\frac{\sqrt{3}-3}{2}} \left(\frac{3+\sqrt{3}}{3-\sqrt{3}}\right)^{\frac{\sqrt{3}-1}{4}}
 (k_{1}r)^{2(1-\sqrt{3})}.
\end{equation}
Using (\ref{k1}) and (\ref{sregulapproxU}) we obtain
\begin{equation}
\frac{P(r)}{r}\approx \frac{r_{M}}{r}-\frac{Q^2}{r^2}
-\alpha_{1}m_{c}^2 r^2\left(\frac{r_{*}}{r}\right)^{2(\sqrt{3}-1)},
\end{equation}
where $\alpha_{1} \approx 0.84$ \cite{gi:plb}.

 The accelerated branch solution is obtained from the solution of (\ref{ktoAccelU}) by imposing the
boundary condition $P(r)+Q^2/r \rightarrow r_{M}$ for $r \rightarrow 0$ and $P(r)-m_{c}^2 r^3
\rightarrow 0$ for large $r$.
In order to determine the integration constant $k_{2}$, imposing the above boundary condition at
the (\ref{P}), we obtain $k_{2}$ from the following relation \cite{gi:plb}
\begin{equation}
\label{k2}
2(k_{2}r_{*})^3=c' \approx 4.41,
\end{equation}
where $c^{\prime}$ is obtained from the following integral
\begin{equation}
c'=\int_{-\infty}^{-2}\left[ - \frac{(- 5 - 3U + f)(-3 - 3U -
       \sqrt{3}f)^{2\sqrt{3}}}{(U + 2)^2( 1 + 3U + f)}\right]^{3/8} dU .
\end{equation}

One can see the following asymptotic behavior by consulting the result of \cite{gi:plb}.

 At large distances, $r \gg r_{*}(U \rightarrow -2^{-})$, (\ref{ktoAccelU}) is approximated as
\begin{equation}
\label{laccelapproxU}
U+2 \approx -\frac{\sqrt{2}(3-\sqrt{3})^{\sqrt{3}}}{2 (k_{2}r)^4}.
\end{equation}
 Using (\ref{k2}) and (\ref{laccelapproxU}) one obtains
\begin{equation}
\frac{P(r)}{r}\approx -\frac{\tilde{r}^2_{M_{2}}+Q^2}{r^2}+m_{c}^2 r^2,
\end{equation}
where $\tilde{r}^2_{M_{2}}= 6^{\sqrt{3}} \frac{3\sqrt{2}}{4(3+\sqrt{3})^{\sqrt{3}}}\frac{m_{c}^2}{k_{2}^4}
=   \frac{3\sqrt{2}}{4(3+\sqrt{3})^{\sqrt{3}}}\frac{m_{c}^2}{k_{1}^4}
\approx 0.56 r_{M}r_{*}$.

As in the regular branch case the screened mass effect occurs, but the screened charge effect does not occur
in the accelerated branch case also.

 At short distances, $r \ll r_{*}(U \rightarrow -\infty)$, since $f \approx -\sqrt{3}(1+U+1/6U)$,
(\ref{ktoAccelU}) is approximated as
\begin{equation}
\label{saccelapproxU}
U \approx -2^{\frac{\sqrt{3}-3}{2}} \left(\frac{3+\sqrt{3}}{3-\sqrt{3}}\right)^{\frac{\sqrt{3}-1}{4}}
 (k_{2}r)^{2(1-\sqrt{3})}.
\end{equation}
Using (\ref{k2}) and (\ref{saccelapproxU}) one obtains
\begin{equation}
\frac{P(r)}{r}\approx \frac{r_{M}}{r}-\frac{Q^2}{r^2}
-\alpha_{2}m_{c}^2 r^2\left(\frac{r_{*}}{r}\right)^{2(\sqrt{3}-1)},
\end{equation}
where $\alpha_{2}=-\alpha_{1} \approx 0.84$ \cite{gi:plb}.
\\

\section{Conclusion}

In this paper we study charged black holes on the brane in
the DGP model. Beginning with an ansatz for the induced metric on
the brane, we solve the constraint equations of
(4+1)-dimensional gravity to find metrics describing charged
brane-world black holes. In the absence of the Maxwell charge, we
obtain the Schwarzschild solution obtained
previously by Gabadadze and Iglesias. In the presence of the Maxwell
charge, we obtain charged black hole solutions which are
the Reissner-Nordstrom type with some corrections. One
type of our solutions exhibits the phenomena of the accelerated
expansion, the same behavior as in the Schwarzschild case. As in
the Schwarzschild case, the screened mass
effect also occurs in the charged case. However, no such effect for the charge appears.
\\



\vspace{2mm}

\noindent
{\large {\bf Acknowledgments}} \\
%
%
\noindent
 This work was supported by the Korea Research Foundation
Grant funded by the Korean Government(MOEHRD),
KRF-2006-312-C00498.
E. C.-Y. thanks KIAS for hospitality during the time that this
work was done.
\\



\end{document}